# Pressure-dependent bandgap study of MBE grown {CdO/MgO} short period SLs using diamond anvil cell

Abinash Adhikari[1,*], Pawel Strak[2], Piotr Dluzewski[1], Agata Kaminska[1,2,3], Ewa Przezdziecka[1]

*[1]Institute of Physics, Polish Academy of Sciences, Al. Lotnikow 32/46, 02-668, Warsaw, Poland*

*[2]Institute of High-Pressure Physics, Polish Academy of Sciences, Sokolowska 29/37, 01-142 Warsaw, Poland*

*[3]Faculty of Mathematics and Natural Sciences. School of Exact Sciences, Cardinal Stefan Wyszynski University, Dewajtis 5, 01-815 Warsaw, Poland*

*Corresponding author: adhikari@ifpan.edu.pl

### Abstract

Semiconductor superlattices have found widespread applications in electronic industries. In this work, short-period superlattice structure (SLs) composed of CdO and MgO layers was grown using the plasma-assisted molecular beam epitaxy (PA-MBE) technique. The optical property of the SLs was investigated by absorption measurement at room temperature. The ambient-pressure direct bandgap was found to be 2.76 eV. The pressure dependence of fundamental bandgap has been studied using a diamond anvil cell (DAC) technique. It has been found that the band-to-band transition shifts toward higher energy with applied pressure. The bandgap of SLs was varied from 2.76 eV to 2.87 eV with applied pressure varied from 0 to 5.9 GPa. The pressure coefficient for the direct bandgap of SLs was found to be 26 meV/GPa. The obtained experimental result was supported by theoretical results obtained using DFT calculations. The volume deformation potential was estimated using the empirical rule. We believe our findings may provide valuable insight for a better understanding of {CdO/MgO} SLs towards their future applications in optoelectronics.

Oxide heterojunctions are presently a subject of strong interest because they provide a wealth of modern functionalities that can be exploited in a broad range of currently emerging technologies involving ultrafast electronics, high-sensitivity chemical sensors, and quantum technology.[1–5] The ability to tailor their properties, in particular, the bandgap, has generated considerable technological interest to use in various device applications.[6,7] SLs have been demonstrated as alternate layers of two different semiconductors to provide a flexible way of bandgap engineering in a wide spectral range.[8] Latterly, advanced developments in epitaxial growth techniques including molecular beam epitaxy (MBE), and metal-organic chemical vapor deposition (MOCVD) allowed us to control the layer thickness precisely and made it possible to design the electronic band structures of semiconductors accordingly for device fabrication.

CdO is one of the oldest known conductive oxides that have been studied widely for the last century despite having a relatively small direct bandgap $E_\Gamma$ ~2.23 eV and two indirect bandgaps $E_L$~1.2 eV and $E_{-X}$~0.8 eV. On the contrary, MgO is a direct bandgap semiconductor having a bandgap of 7.5 eV and exists in the same as CdO, cubic rocksalt ($F\bar{m}3m$) structure. It is worth noting that, despite having an isostructural configuration and possible bandgap tunning in a wide spectral range, the study on (Mg, Cd)O ternary alloys is limited.[9–12] Przezdziecka *et al.* reported quasi ternary alloys short period {CdO/MgO} SLs grown on a sapphire substrate using the MBE technique[13] that demonstrated, the variation of bandgap energy can be modulated in a wide range from 2.6 eV to 6 eV by changing the CdO sublattices thickness while keeping the same thickness of MgO sublatices.[14]

The pioneering work by Charlie Weir and Alvin Van Valkenburg who made the first diamond anvil cell (DAC),[15] initiated the research on high-pressure study in many experimental techniques.[16–19] Applying pressure leads to a change in atomic volume in the materials which affects the electronic band structure of the semiconductor materials, and hence the bandgap. For bulk semiconductor materials, the change in bandgap is straightforward as one can determine whether the optical transition across the bandgap is direct or indirect in k-space. However, for SLs the position of valence-band maximum (VBM) and conduction-band minimum (CBM) can be very close to SL layer (spatially direct) or it can be at a different location (indirect in real space).[20] Hence, determining the bandgap is rather complicated. Previously, the variation of bandgap with applied hydrostatic pressure was determined for bulk AlN,[21] GaN,[22] InN,[23] and ZnO[24,25] which were in reasonable agreement with the theoretical calculations.[21,26,27] On the other hand, there has been limited available theoretical data[28–30] on the pressure dependence of energy bandgap of SLs with very little experimental evidence.[31,32] This fact motivated us to study the dependence of the fundamental bandgap of {CdO/MgO} SLs with applied hydrostatic pressure using DAC. In this letter, a short period {CdO/MgO} SLs was grown on an *r*-plane (1-102) sapphire substrate by plasma-assisted MBE technique using a Riber Compact 21B system (supplementary material). To provide a better understanding of bandgap variation with applied hydrostatic pressure in SLs, we have performed optical absorption measurements using DAC. The pressure coefficient (PC) was determined and compared with the result obtained from theoretical calculations (supplementary material). The effective volume deformation potential of SLs was also determined. Together with theoretical analysis, our data provide an overall picture of the hydrostatic pressure behavior of the fundamental bandgap of {CdO/MgO} SLs.

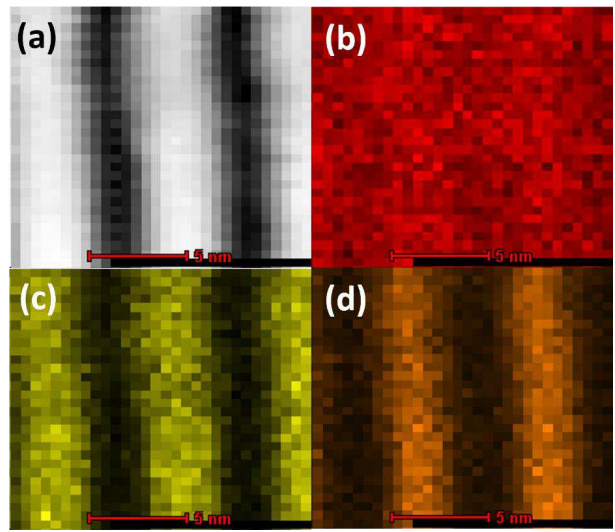

Fig. 1. (a) STEM/HAADF image of CdO/MgO superlattice and EDX mapping of (b) oxygen, (c) cadmium and, (d) magnesium.

High-resolution transmission electron microscopy (HR-TEM) image and elemental composition maps using EDX technique confirmed the separation of Mg and Cd elements and distinct interface between CdO and MgO layers as shown in Fig.1. The ambient-pressure room temperature transmittance spectra of short-period {CdO/MgO}$_{21}$ SLs was determined using UV-Vis spectrometer. The bandgap was determined from Tauc's relation (supplementary material).

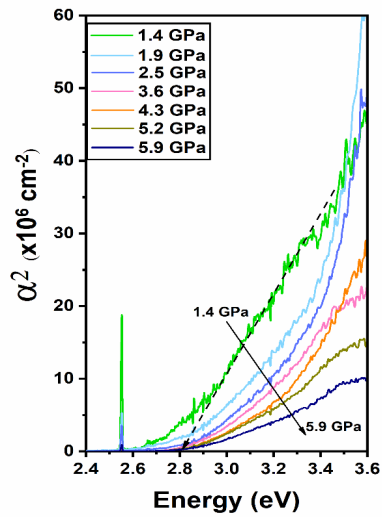

Fig.2. $\alpha^2$ plot as a function of energy (hν) for {CdO/MgO} SLs at different hydrostatic pressure.

The pressure-dependent fundamental bandgap was determined using absorption spectroscopy with the sample placed inside a DAC. From the transmittance spectra, the direct bandgap was determined using Tauc's relation with n value equal to 1/2. Figure 2 shows a variation of the square of the absorption coefficient (α) of SLs vs. photon energy with a change in applied hydrostatic pressure in the DAC. With the increase of hydrostatic pressure in the DAC, the absorption edge shifts towards higher energy, implying an increase of the bandgap of SLs. It is worth noting that, because of a relatively small aperture window in the DAC along with a low signal-to-noise ratio, it is difficult to obtain the precise transmittance behavior. And hence we have obtained noisy absorption coefficient data (as shown in Fig. 2). The bandgap can be determined by carefully extrapolating the linear portion of $\alpha^2$ to the point of intersection with the energy axis. This extrapolation procedure provides a strong and consistent method to determine the pressure dependence of direct bandgap (Γ- Γ transition). In CdO indirect band gaps are present at L and around X point. Thus, the transition in SLs can be mixed of direct and indirect characters. Owing to the complexity of the calculations from obtained absorption data, the indirect bandgaps (Γ-X and Γ-L transition) have not been taken into consideration. The pressure dependence of the direct bandgap of SLs is plotted in Fig. 3. The error bars of the bandgap originate from the uncertainty in the extrapolation of absorption data down to the energy axis. It is observed that the direct bandgap shifts to higher energy from 2.76 eV to over 2.87 eV with applied pressure varied from 0 to 5.9 GPa in DAC. Wei and Zunger,[33] and Shan and Walukiewicz *et al.*[34] corrected the fitting equation to liner pressure-dependent function:

$$E_g(P) = E_g(0) + \alpha_P P \qquad (1)$$

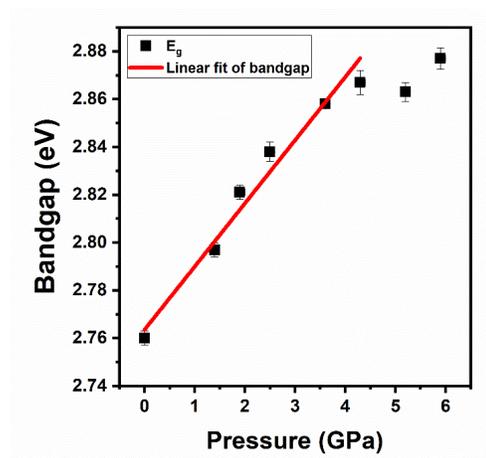

Fig 3. Variation of bandgap energy as a function of applied hydrostatic pressure

where $\alpha_P = dE_g/dP$, termed as liner pressure co-efficient (PC). The obtained experimental data are fitted using equation 1. The zero-pressure bandgap energy ($E_g(0)$) and the PC of the direct bandgap of SLs found to be (2.761±0.002) eV and (26±2) meV/GPa respectively. The calculated zero pressure bandgap energy agrees with the experimental data obtained using UV-Vis spectroscopy at ambient conditions within ~1% error. It is observed that the optical bandgap increases linearly as the pressure in DAC increases up to 4.3 GPa which is attributed to direct transition (Γ- Γ transition). At higher pressures, the change in optical bandgap with pressure changes nonlinearly. In general, with applied hydrostatic pressure, the wavefunctions of neighboring atomic orbitals of individual Cd and Mg start to overlap. Therefore, the band transition comes probably from the mixed contribution of direct and indirect transitions in SLs which have different PCs.[35] The pressure in the DAC is high enough to achieve Γ- X and Γ- L crossover along with Γ- Γ transition. The non-linear bandgap energy behavior above 4.3 GPa can also be attributed to non-hydrostatic effects at higher pressures connected with the extremely sophisticated double-gasket measuring technique of transmittance spectra (Supplementary Material). In spite of stable line width of ruby R1 line in this range of pressure it could happen, that the compressed gasket started to exert some kind of axial stress on the sample, which is much larger than the ruby ball, so the pressure applied to the sample can be different than the pressure indicated by ruby luminescence encountered in DAC measurements. It is worth noting that, the exact analysis of bandgap behavior with applied pressure in the DAC is complex. A small variation in intensity or experimental setup caused e.g., by changes in the sample chamber size with increasing pressure can lead to problems with the interpretation of the results. In order to get more information about the behavior of bandgap along with indirect transition in the investigated SLs, we have performed theoretical calculations using VASP code which is described in the following section.

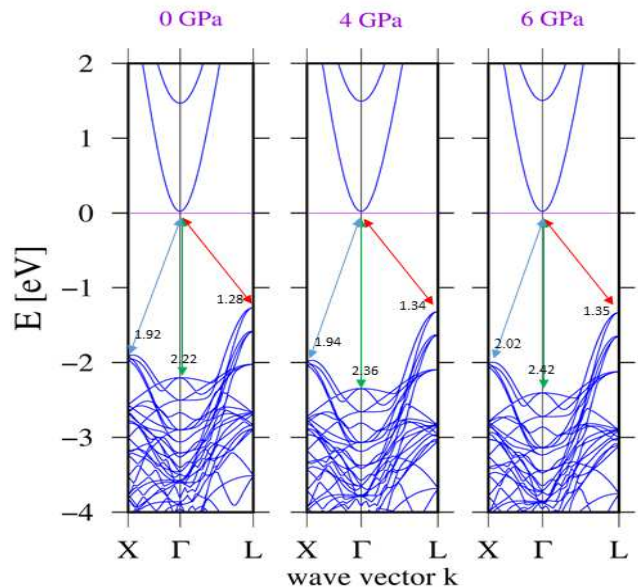

Fig 4. Calculated electronic band structure of 4ML {CdO/MgO} SLs at (a) ambient condition (0 GPa), (b) 4 GPa, and (c) 6 GPa.

The Perdew, Burke, and Ernzerhof (PBE) density functional approach provides incorrect values for bandgaps of semiconductors. Several methods have been used to remove this deficiency, such as an approximation to the self-energy (GW approximation),[36] hybrid functionals using Hartree–Fock correction,[37] or half-occupation generalized-gradient approximation (GGA-1/2).[38] In the reported calculation we used the most efficient latter scheme, proposed by Ferreira et al.[38] Spin–orbit effects were neglected since the high-lying valence states and low-lying conduction states lead to a small splitting (of the order of 10 meV). The calculated bandgaps of bulk MgO and CdO were $E_\Gamma(MgO) = 7.1$ eV and $E_\Gamma(CdO) = 2.55$ eV, respectively which are in good agreement with experimental results obtained in the literature.[39–41] This completes the above-mentioned second stage in which the final results are obtained by application of the modified GGA-1/2 correction method to structures in which the positions of atoms and periodic cell size were determined in the first stage using the PBE approximation. It is seen that PBE underestimates the value of the energy gap, while in GGA-1/2 it is calculated correctly. The band structures of {CdO/MgO} SLs at 0, 4 and 6 GPa pressure are shown in Fig. 4. Hydrostatic pressure dependences of the bandgap energies of three special points in the Brillouin zone giving the PC are shown in Fig. 5. From that data it follows that Bandgap change is different in a different point of the Brillouin zone.

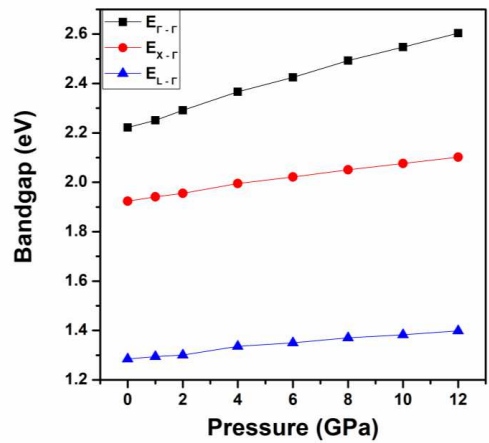

Fig.5. Calculated variations of 4ML {CdO/MgO} SLs bandgaps with applied pressure.

The linear PC ($\alpha_P$) of CdO, MgO (from the literature), and {CdO/MgO} SLs determined from experiment and theoretical calculation are listed in Table1. The obtained experimental value of PC is lower compared to the value obtained from theoretical calculations. In-plane tensile strain and out-of-plane compressive strain of the MgO and CdO layers is expected in SLs and their values depend on sublayers thickness.[14,42] In the case of the experimental procedure, the SLs was grown on the *r*-plane sapphire substrate which could lead to a lattice mismatch condition, and hence a residual strain in the SLs arises. Furthermore, sapphire has a larger bulk modulus (~ 250 GPa) compared to CdO (~ 140 GPa) and MgO (~ 170 GPa),[43] which causes the SLs experiences a lower effective pressure than the pressure applied in the DAC.[44] Therefore, due to substrate and the strain effect, the obtained experimental value of PC is lower compared to the calculated theoretical one. The positive value of PC implies a bandgap widening with the application of hydrostatic pressure. It is found that the PC of SLs corresponding to $\Gamma$- $\Gamma$ transition is higher compared to $\Gamma$- L and $\Gamma$- X transition which suggests a profound direct transition compared to indirect transitions in SLs.

It is seen that the bandgap PC of {CdO/MgO} SLs, falls in-between the bandgap PC of CdO and MgO (Table 1). From the reported values of bandgap PC of CdO and MgO in the literature as mentioned in the Table 1, the following explanations can be withdrawn.

(i) In the case of CdO, the more favorable transition is for smaller bandgap i.e $\Gamma$- X transition, and the PC can lead to direct-indirect bandgap transition. With further increase of pressure, the phase of CdO structure is distorted and changes to a stable (Zinc-blende or wurtzite) structure. Previously, Sahoo *et al.* performed ab initio calculation and reported that the phase transition of CdO from B1 (rocksalt type) → B2 (CsCl type) occurs at a pressure of about 87 GPa.[45]

(ii) MgO is a direct bandgap semiconductor stable in cubic rocksalt structure at ambient conditions in which the PC corresponds to direct transition ($\Gamma$- $\Gamma$). It requires a very high pressure of 509 GPa for phase transition from B1→ B2 structure.[46] Another theoretical calculation of MgO suggests that a high-pressure value of 515 GPa is required for the phase transition of MgO.[47]

(iii) The PC of CdO is lower than that of MgO. Wei and Zunger[33] reported that the bandgap PC decreases with an increase in cation atomic number (from Mg to Cd) in the same anionic system. According to their theory, the PC is influenced by *p-d* coupling in the material. In the case of CdO, Cd has a deeper *3d* state which contributes to *p-d* coupling, hence a decrease in PC is observed. However, for MgO, it lacks an occupied *d* orbital in Mg and a higher PC is obtained compared to CdO.[48] Similar behavior has been reported in the literature which suggests a decrease in bandgap PC with an increase in the ionicity of the materials.[49] However, the physical origin of this behavior in the case of a common anion system is rather complicated and can be explained using *s-s, p-p, p-d* coupling along with the bulk modulus of the materials.

**Table 1** Experimental and calculated pressure co-efficient ($\alpha_P$, in meV/GPa) of CdO, MgO, {CdO/MgO} SLs

|  | $\alpha_P^{\Gamma-\Gamma}$ (meV/GPa) | $\alpha_P^{\Gamma-L}$ (meV/GPa) | $\alpha_P^{\Gamma-X}$ (meV/GPa) |
|---|---|---|---|
| **{CdO/MgO} SLs** | 26[a] | | |
| | 32[b] | 14.9[b] | 9[b] |
| **CdO** | 39.6[c] | 27.3[c] | 50.1[c] |
| | 46.3[d] | 35.1[d] | 62.7[d] |
| | 15[e] | | |
| **MgO** | 57.9[c] | 36.3[c] | 5.4[c] |
| | 64.8[d] | 41[d] | 10[d] |
| | 37.3[f] | | |

[a]This work (exp), [b]This work (theory), [c]Ref.[50], [d]Ref.[51], [e]Ref.[52], [f]Ref.[53]

A consequence of hydrostatic pressure in SLs is a change in lattice parameters of constituent structure and hence, a change in volume of the SLs occurs. The relative change in volume of SLs induced by pressure can be calculated using Murnaghan equation of state,[54]

$$P(V) = \frac{B}{B'}\left[\left(\frac{V}{V_0}\right)^{-B'} - 1\right] \quad (2)$$

where $B$ is the bulk modulus, $B'$ is the pressure derivative of bulk modulus ($B' = dB/dP$), $V_0$ is the volume at ambient condition, and V is the change in volume induced by pressure. The results obtained from the DFT calculation, are used here to determine the bulk modulus of the SLs. The data is optimized at total minimum energy at their equilibrium volume. $B$ and $B'$ are calculated by fitting the Pressure-Volume data using the relations derived from Murnaghan equation of state, $B = (-dP)/(d\, lnV)$, and $B' = dB/dP$ respectively. $B$ and $B'$ values of SLs are found to be 156.6 GPa and 4.75 respectively. For a better comparison, the value of $B$ and $B'$ of SLs (this work) along with CdO and MgO (previously reported in the literature) are listed in Table 2. It is observed that, the reported value of the bulk modulus of MgO is higher compared to that of CdO. The variation of reported data comes from different theoretical approaches to calculate $B$ and $B'$.

**Table 2** Calculated bulk modulus (B in GPa), pressure derivative of bulk modulus $B'$, and volume deformation potential ($α_V$ in eV)

|  | $B$ (GPa) | $B'$ | $α_V$ (eV) |
|---|---|---|---|
| **{CdO/MgO} SLs** | 156.6[a] | 4.75[a] | -5.05[a] |
| **CdO** | 166[b] | 5.08[b] | -10.08[b] |
|  | 143[c] | 4.71[c] | -6.65[c] |
|  | 130[d] | 4.13[d] |  |
| **MgO** | 174[b] | 4.24[b] | -6.35[b] |
|  | 179[c] | 4.14[c] | -11.63[c] |
|  | 171[e] | 4.29[e] |  |
|  | 186[f] | 3.53[f] |  |

[a]This work, [b]Ref.[50], [c]Ref.[51], [d]Ref.[55], [e]Ref.[56], [f]Ref.[57],

Due to the difference in lattice parameters of CdO and MgO, strain is present in between the sublayers of SLs. Moreover, the application of hydrostatic pressure to SLs causes an additional strain that results in a volume change, and hence as a consequence, there is a shift of the conduction band edge relative to the valence band edge. Under hydrostatic pressure conditions, the linear displacement coefficient is defined as volume deformation potential ($α_V$) which is essential for band offset calculations in heterostructure device applications. It also describes the scattering mechanism of semiconductors assuming that the local lattice deformation produced by the phonons is equivalent to homogenous deformation in the crystals. The experimental procedure of volume deformation potential determination is quite difficult as one cannot determine all the strain components of an SLs, and the absolute shift of the band edge accurately under hydrostatic pressure conditions. In this work, we have determined the hydrostatic volume deformation potential of SLs by fitting the bandgap energy and volume data obtained from a theoretical calculation using a relation

$$α_V = \frac{dE_g}{d\ lnV} \qquad (3)$$

The volume deformation potential of SLs is found to be -5.05 eV. In this calculation, the bandgap corresponding to direct transition (Γ- Γ) has been taken into consideration. From Murnaghan equation of state, the value of bandgap PC ($α_P$) and volume deformation potential ($α_V$) of SLs are related through bulk modulus (*B*) using relation, $α_P = - α_V/B$. For a better comparison, previously reported volume deformation potentials of CdO and MgO are listed in Table 2. For all the cases, $α_V$ was found to be negative. The negative sign arises from the kinetic energy effect and a strong anion *s* – cation *s* repulsion. Previously it is reported that, the $α_V$ of CdO is higher compared to MgO (i.e. less negative) because of an increase in cation-anion bond length with increases in cation atomic number for the same anionic system.[50] Application of hydrostatic pressure to a solid body causes its contraction: in elastic (reversible) and plastic (irreversible) regimes. In the ab initio simulations, the complexity of plastic deformation is usually drastically reduced by an approximation of the change in obtained lattice parameters to linear and quadratic pressure terms. No irreversible deformation is therefore taken into account. In the performed investigations, the structure was limited to reversible changes, i.e., generation and motion of dislocations were not considered. Naturally, this assumption limits the application of the results to moderately strained structures. In high-pressure experiments, strain alters the interatomic distances and hence the lattice parameters of CdO and MgO change, and the bandgap is affected. Therefore, the changes in lattice parameters are the basic variables describing the influence of pressure on {CdO/MgO} SLs strained systems.

In conclusion, the results obtained here describe the dependence of optical bandgap variation in {CdO/MgO} SLs with applied hydrostatic pressure. The bandgap PC was determined from absorption spectroscopy using a DAC and found to be 26 meV/GPa. Using GGA-1/2 method within the framework of DFT formalism, the electronic band structure of {CdO/MgO} SLs was determined. It is found that the Γ- Γ, Γ- L, and Γ- X bandgap increases with an increase in pressure, however, the Γ- Γ transition dominates over the other two indirect transitions. A higher PC of value 32 meV/GPa was determined from the theoretical calculation for ideal SLs. The bulk modulus and the derivative of bulk modulus of the {CdO/MgO} SLs are reported and compared with the existing data for CdO and MgO in the literature. The bandgap deformation potential of SLs was determined to be -5.05 eV. We believe that this measurement technique using DAC along with the theoretical calculation will provide an insight into the characteristics of short-period oxide SLs with relevance to "bandgap engineering" and consequently apply to device designing and optimization, e.g., pressure sensors.

See the supplementary material for details of the growth of {CdO/MgO} SLs using plasma-assisted MBE technique. The experimental setup for bandgap measurements is described in detail. We also provide a room temperature bandgap determination procedure using Tauc relation.

This work was supported by the Polish National Science Center Grant No. 2021/41/N/ST5/00812, and 2021/41/B/ST5/00216. This research was carried out with the support of the Interdisciplinary Centre for Mathematical and Computational Modelling at the University of Warsaw (ICM UW) under grants no GB84-23.

**AUTHOR DECLARATIONS**

**Conflict of Interest**

The authors have no conflicts to disclose.

**Author Contributions**

**Abinash Adhikari:** Data curation (equal); Formal analysis (lead); Investigation (lead), Methodology (equal); Writing-Original draft (lead); Funding Acquisition (equal)**. Pawel Strak:** Data curation (equal); Formal analysis (supporting); Writing-Original draft (supporting); Software (lead); **Piotr Dluzewski:** Investigation (supporting), **Agata Kaminska:** Conceptualization (equal)**;** Methodology (equal); Resources (equal); Supervision (equal); Review & Editing (equal)**.** **Ewa Przezdziecka:** Conceptualization (equal), Resources (equal); Funding Acquisition (equal); Project administration (lead), Supervision (equal), Review & Editing (equal)

**Data availability**

The data that support the findings of this study are available from the corresponding author upon reasonable request.

ACCEPTED MANUSCRIPT
Applied Physics Letters
AIP Publishing
This is the author's peer reviewed, accepted manuscript. However, the online version of record will be different from this version once it has been copyedited and typeset.
PLEASE CITE THIS ARTICLE AS DOI: 10.1063/5.0123342Growth Des. **17**, 6303 (2017).

[12] A. Adhikari, A. Lysak, A. Wierzbicka, P. Sybilski, A. Reszka, B.S. Witkowski, and E. Przezdziecka, Mater. Sci. Semicond. Process. **144**, 106608 (2022).

[13] E. Przezdziecka, A. Wierzbicka, P. Dłuzewski, I. Sankowska, P. Sybilski, K. Morawiec, M.A. Pietrzyk, and A. Kozanecki, Cryst. Growth Des. **20**, 5466 (2020).

[14] E. Przeździecka, P. Strąk, A. Wierzbicka, A. Adhikari, A. Lysak, P. Sybilski, J.M. Sajkowski, A. Seweryn, and A. Kozanecki, Nanoscale Res. Lett. **16**, 59 (2021).

[15] C.E. Weir, E.R. Lippincott, A. Van Valkenburg, and E.N. Bunting, J. Res. Natl. Bur. Stand. Sect. A Phys. Chem. **63A**, 55 (1959).

[16] T. Kobayashi, Rev. Sci. Instrum. **56**, 255 (1998).

[17] A. Jayaraman, Rev. Sci. Instrum. **57**, 1013 (1986).

[18] G. Shen and H.K. Mao, Reports Prog. Phys. **80**, 016101 (2016).

[19] A. Jayaraman, Rev. Mod. Phys. **55**, 65 (1983).

[20] I. Gorczyca, T. Suski, N.E. Christensen, and A. Svane, Appl. Phys. Lett. **101**, 1 (2012).

[21] H. Akamaru, A. Onodera, T. Endo, and O. Mishima, J. Phys. Chem. Solids **63**, 887 (2002).

[22] P. Perlin, L. Mattos, N.A. Shapiro, J. Kruger, W.S. Wong, T. Sands, N.W. Cheung, and E.R. Weber, J. Appl. Phys. **85**, 2385 (1999).

[23] S.X. Li, J. Wu, E.E. Haller, W. Walukiewicz, W. Shan, H. Lu, and W.J. Schaff, Appl. Phys. Lett. **83**, 4963 (2003).

[24] A. Kaminska, A. Duzynska, M. Nowakowska, A. Suchocki, T.A. Wassner, B. Laumer, and M. Eickhoff, J. Alloys Compd. **672**, 125 (2016).

[25] A. Duzynska, R. Hrubiak, V. Drozd, H. Teisseyre, W. Paszkowicz, A. Reszka, A. Kaminska, S. Saxena, J.D. Fidelus, J. Grabis, C.J. Monty, and A. Suchocki, High Press. Res. **32**, 354 (2012).

[26] I. Gorczyca, T. Suski, N.E. Christensen, and A. Svane, Phys. Rev. B **83**, 153301 (2011).

[27] Y. Duan, L. Qin, L. Shi, and G. Tang, Comput. Mater. Sci. **101**, 56 (2015).
14

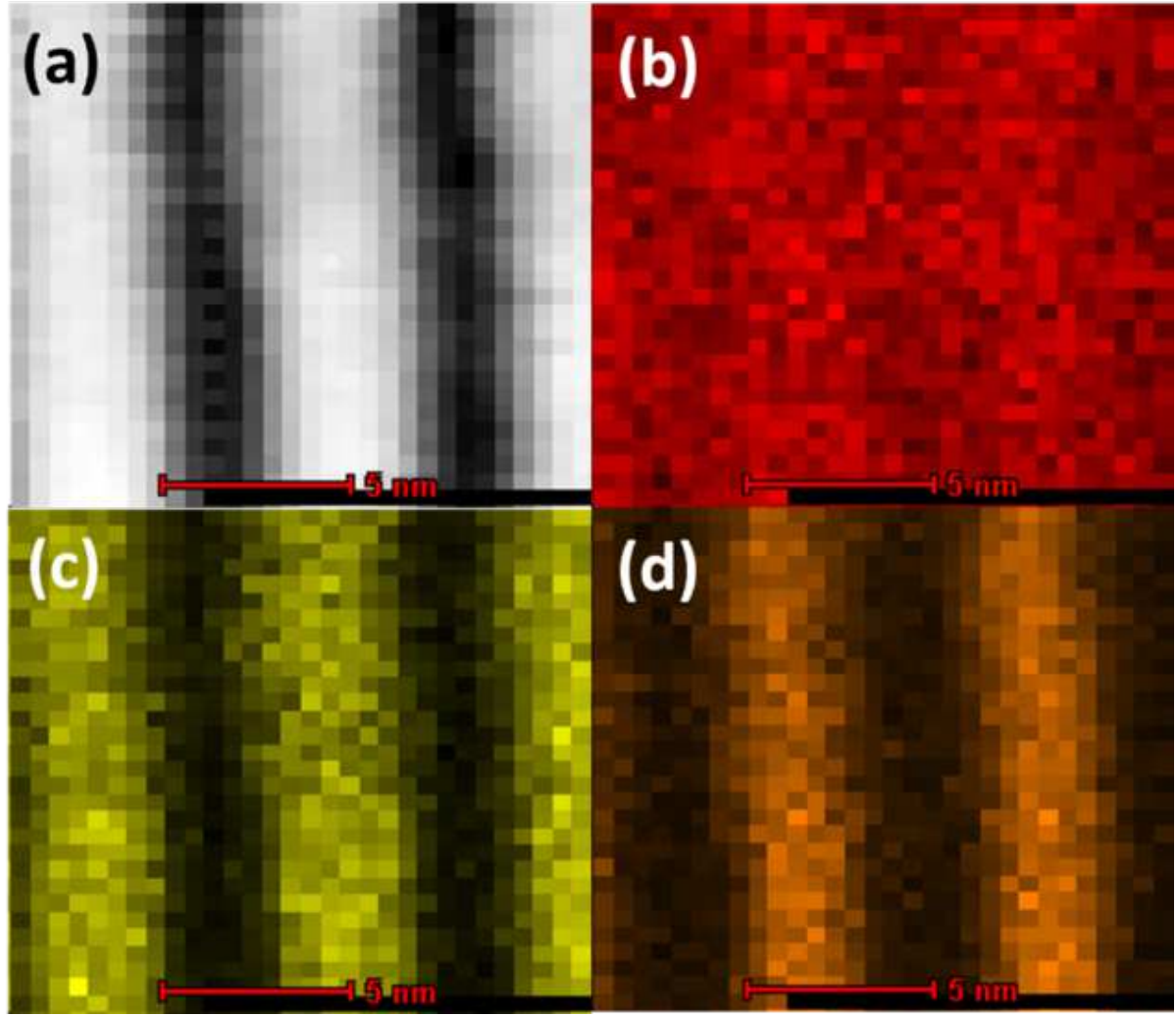

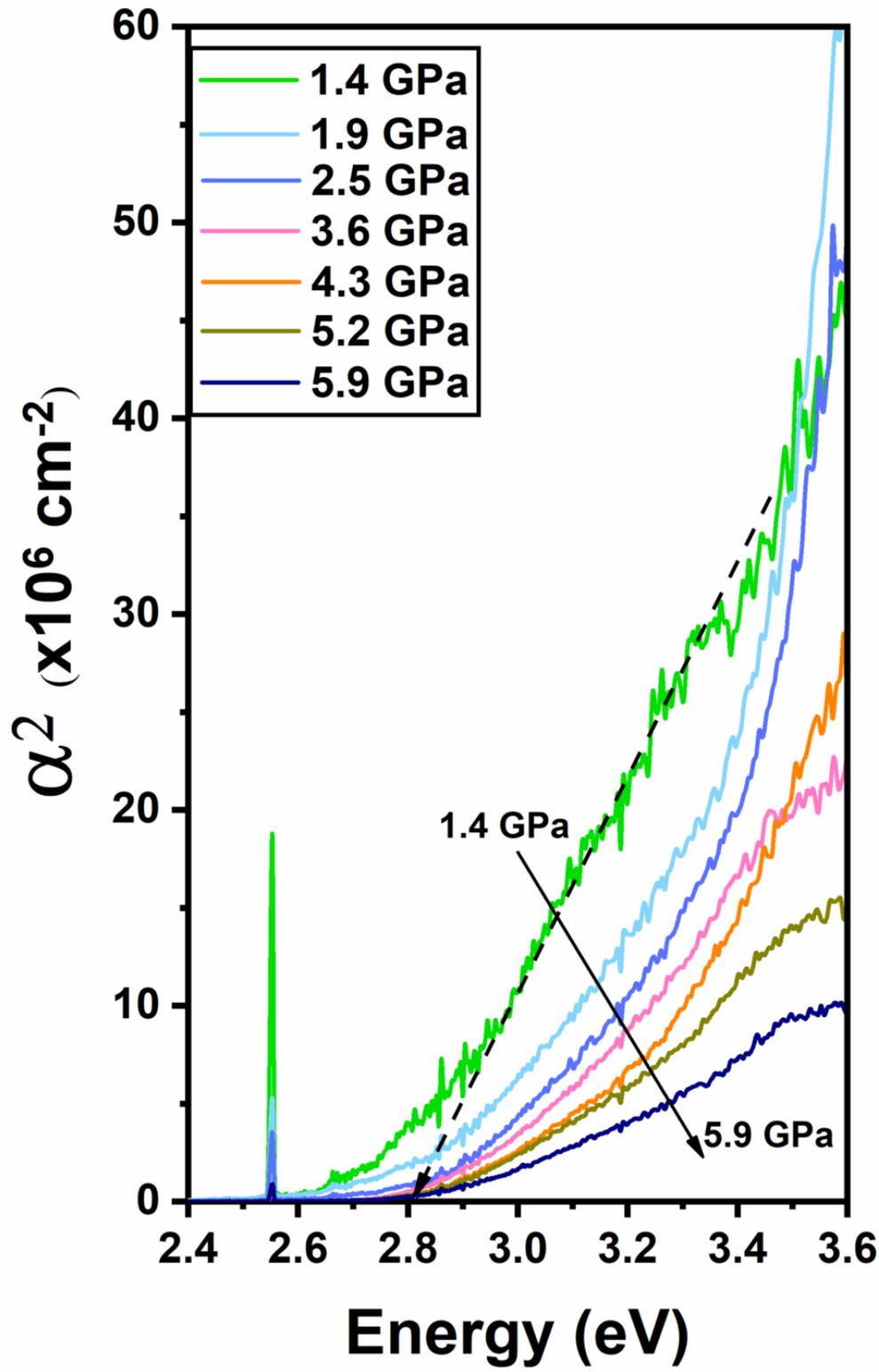

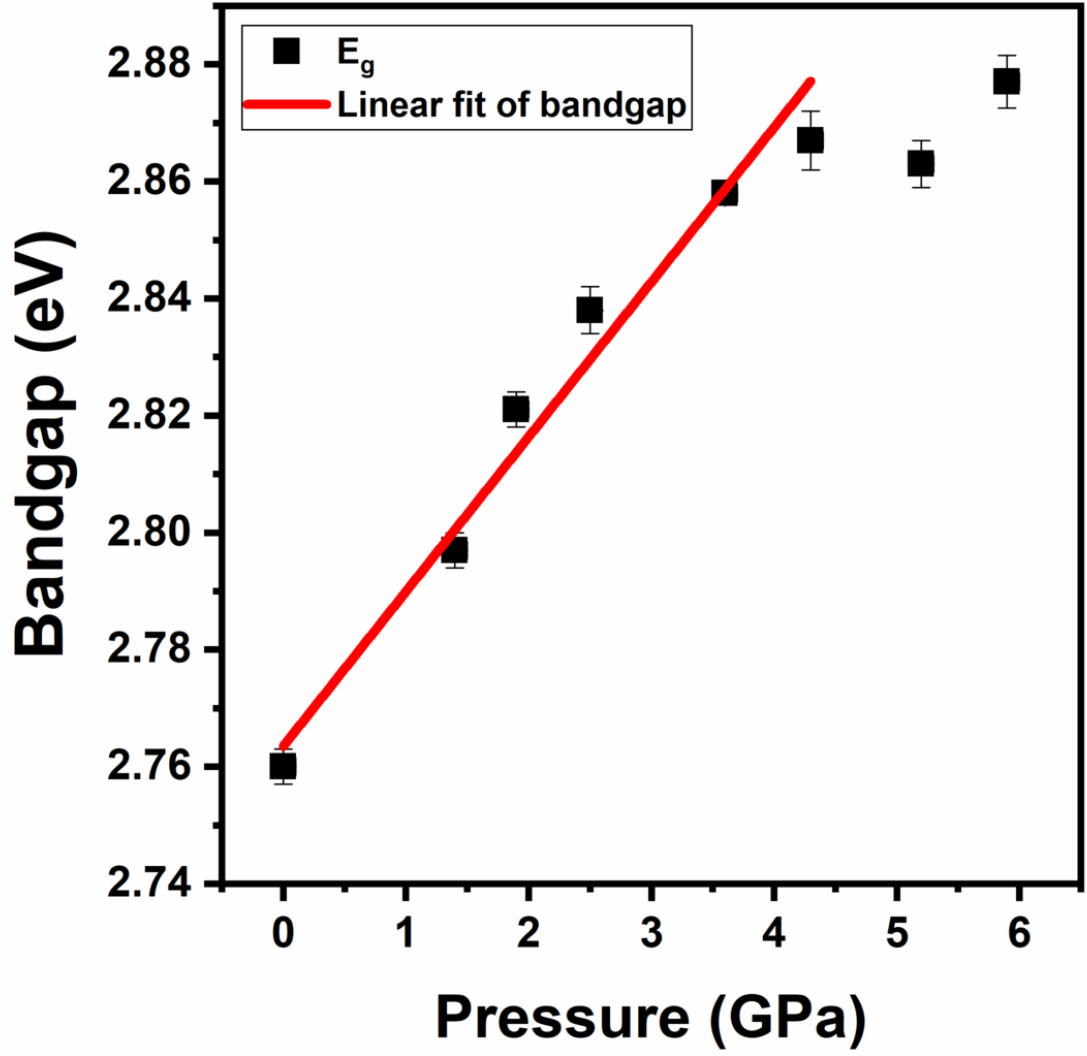

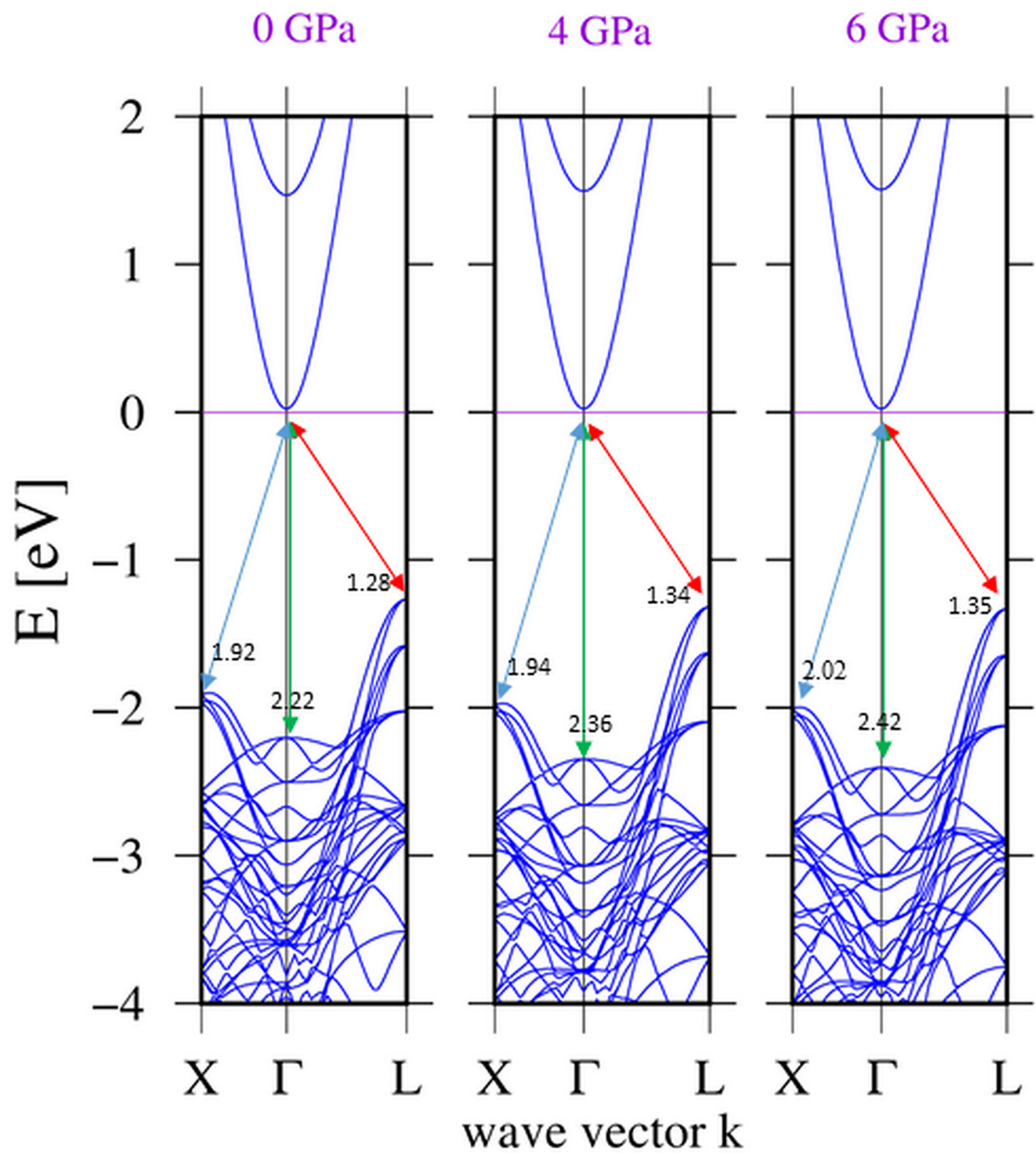

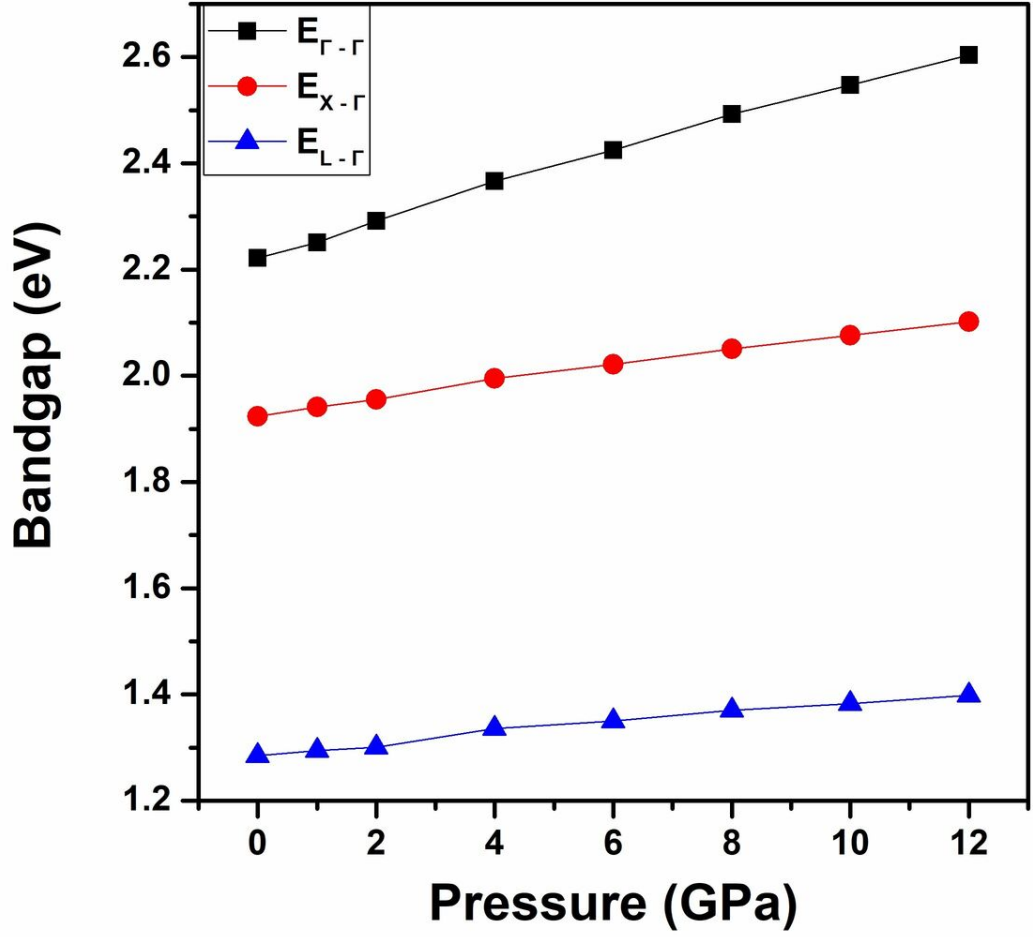